\documentclass[submission,copyright,creativecommons]{eptcs}
\providecommand{\event}{ACL2 2014} 

\usepackage{amsmath}
\usepackage{amssymb}
\usepackage{latexsym}
\usepackage{mathpartir}
\usepackage{multicol}
\usepackage{alltt}
\usepackage{url}

\newcommand{\ie}{\emph{i.e.}, }
\newcommand{\eg}{\emph{e.g.}, }

\newcommand{\Ra}{\Rightarrow}

\renewcommand{\And}{\wedge}

\newcommand{\subtype}[2]{#1 \sqsubseteq #2}

\newcommand{\CH}{\ensuremath{\mathcal{H}}}

\renewcommand{\phi}{\varphi}

\newcommand{\oneof}{\textsf{oneof}}
\newcommand{\all}{\textsf{all}}

\newcommand{\p}[1]{\begin{normalfont}\frenchspacing\texttt{#1}\end{normalfont}}
\newcommand{\B}{\ | \ }

\newcommand{\true}{\ensuremath{\mathit{true}}}

\title{Data Definitions in the ACL2 Sedan}
\author{Harsh Raju Chamarthi \quad\quad Peter C. Dillinger \quad\quad Panagiotis Manolios\institute{College of Computer and Information Science}
\institute{Northeastern Univeristy}
\email{harshrc@ccs.neu.edu \quad\quad pcd@ccs.neu.edu \quad\qquad pete@ccs.neu.edu}
}
\def\titlerunning{Data Definitions in ACL2 Sedan}
\def\authorrunning{H.R. Chamarthi, P.C. Dillinger \& P. Manolios}
\begin{document}
\maketitle

\begin{abstract}
We present a data definition framework that enables the convenient
specification of data types in ACL2s, the ACL2 Sedan.  Our primary
motivation for developing the data definition framework was
pedagogical. We were teaching undergraduate students how to reason
about programs using ACL2s and wanted to provide them with an
effective method for defining, testing, and reasoning about data types
in the context of an untyped theorem prover. Our framework is now
routinely used not only for pedagogical purposes, but also by advanced
users.

Our framework concisely supports common data definition patterns, \eg
list types, map types, and record types. It also provides support for
polymorphic functions.  A distinguishing feature of our approach is
that we maintain both a predicative and an enumerative
characterization of data definitions.

In this paper we present our data definition framework via a sequence
of examples.  We give a complete characterization in terms of tau
rules of the inclusion/exclusion relations a data definition induces,
under suitable restrictions.  The data definition framework is a key
component of counterexample generation support in ACL2s, but can be
independently used in ACL2, and is available as a community book.
\end{abstract}

\section{Introduction}
Data definitions are an essential part of crafting programs and
modeling systems. Whereas most programming languages provide rich
mechanisms for defining datatypes, ACL2 only really provides a
limited collection of built-in types and \p{cons}~\cite{car}. 

This state of affairs presented us with a challenge when we
started teaching undergraduates how to model, specify and reason
about computation, because even freshmen students have a
type-centric view of the world. This led to us squandering class
time, a very limited resource, teaching students how to encode
types using \p{cons}, how to debug such encodings, and how to
reason about them.

We introduced the \p{defdata} framework in ACL2s in order to
provide a convenient, intuitive way to specify data definitions.
A version of \p{defdata} has appeared in ACL2s since at least August
2009 (in version 0.9.7), and we have been extending and improving
it since then. 

Data definitions are critical to how we currently teach students
to model, specify and reason about computation.  They provide
\emph{recursion templates} that students use to define recursive
functions over datatypes.  Students define functions using the
ACL2s macro \p{defunc}, which supports function definitions with
input and output contracts. In contrast to guards, \p{defunc}'s
input contracts allow users to specify the input and output types
of functions in a way that affects the logical meaning of
function definitions.  Data definitions also provide induction
schemes that are used to reason about such functions.

Specifying properties that programs and models satisfy is an art
that takes time to learn. One way of helping beginners master
this skill is to provide them with counterexamples to their
conjectures quickly. Here too, data definitions play a key role
because almost all specifications include hypotheses specifying
the datatypes of the variables. Finding a counterexample to a
conjecture requires satisfying the hypotheses, which requires
finding instantiations of the variables that satisfy their data
definitions. Our \p{defdata} framework maintains both a
predicative characterization, via a predicate recognizing
elements of the datatype, and an enumerative characterization,
via a function that can be used to enumerate all the elements of
the datatype. ACL2s picks out the recognizers in a conjecture and
can use the datatype enumerators to generate ``random'' elements
of the datatype. This is a key part of counterexample generation
in ACL2s~\cite{acl2-11,chamarthi2011automated}. We have found
that the automatic generation of counterexamples to invalid
student conjectures (the common case) is a very effective way of
training students to correctly specify properties of programs
and models. This training happens whenever they use ACL2s, not
just during class. 

The \p{defdata} framework also allows us to increase the amount
of automation ACL2s provides for reasoning about data
definitions.  This increase in automation allows us to reclaim
class time and to use it for more interesting topics.  For
example, our framework generates useful theorems, with
appropriate rule-classes, for list types; it generates accessor
and constructor functions for records with a suitable theory for
reasoning about compositions of these functions; it generates
theorems that characterize the type relations such as inclusion
and exclusion; and it generates events that support polymorphic
type reasoning. 

While the original motivation was pedagogical, we now routinely
use the \p{defdata} framework in our work, as do other advanced
users.  In order to make \p{defdata} more widely available, we
have released it as a community book. This makes it very easy for
regular ACL2 users to take advantage of \p{defdata}.

The paper is organized as follows. We present a number of
examples illustrating the use of the \p{defdata} form in
Section~\ref{sec:eg}; the section is detailed enough to serve as
a rough user guide. We describe the syntax and semantics of the
\p{defdata} language in Section~\ref{sec:full-defdata}. We show
how to characterize the type relations induced by a \p{defdata}
command in Section~\ref{sec:tau-characterize}. We show how
polymorphic functions are supported within our framework in
Section~\ref{sec:poly}. We compare with related work in
Section~\ref{sec:rel} and conclude in Section~\ref{sec:conc}.

\section{Defdata -- Usage and Examples}
\label{sec:eg}
The \p{defdata} macro is the primary method for defining
a new data type.\footnote{In ACL2 one cannot add new objects to
  the universe, one can only partition the existing universe in new
  ways.}  It provides a convenient and intuitive
language to specify type combinations. The \p{defdata} macro names
certain ``type expressions'' whose syntax will be evident from
examples below. A precise description of defdata type expressions and
the defdata form appears in Section~\ref{sec:full-defdata}. It
suffices for now to assume that a defdata form is an event whose
syntax is (\p{defdata} $M$ \emph{body}), where $M$ is a symbol and
\emph{body} a type expression (usually representing some type
combination). When submitted, the event introduces a new typename
($M$), predicate and enumerator definitions for the type, and a host
of other events that support a ``typed'' language setup.

We call $M$ a \emph{defdata type} if it is a name (symbol) that has
been introduced by a \p{defdata} event or is a primitive or custom
type \emph{registered}\footnote{Using the \p{register-type}
  macro, as
  is explained in Section~\ref{sec:full-defdata}.} by the user.

We present a running example to showcase most of the features of
\p{defdata}. Let us suppose we are doing systems-level
modeling. We will see how parts of a processor, filesystem,
etc. can be specified in the \p{defdata} language. The reader, if
she so wishes, can also submit the following commands in an ACL2
session, but the first two events should be skipped in an ACL2
Sedan session.

\begin{alltt}
(include-book "cgen/defdata" :dir :system)
(include-book "cgen/base" :dir :system)
\end{alltt}

The first event sets up the data definition framework (defines
\p{defdata} and other macros). The second causes all primitive types
in ACL2 to be preregistered as defdata types: this means you can use
them in the body of a \p{defdata} form. A good question the user might
ask now: how does one refer or use a primitive type, in other words,
how do we find the typename corresponding to a recognizer/predicate?
Following the ACL2 (and Common Lisp) convention, we try to stick to
the following syntactic rule of thumb. Each \emph{typename} is the
symbol obtained after dropping suffix ``p'' from the predicate name,
and vice versa, \ie given typename \p{M}, the predicate name, \p{Mp},
is obtained by adding suffix ``p''\footnote{An exception is \p{atom};
  it shares the same name as its predicate name. In general, the
  typename and its recognizer need not be related syntactically.}.

\begin{alltt}
(defdata inode nat)
\end{alltt}
In the above example, we created an alias type for natural numbers
whose recognizer is \p{natp}; quite naturally, we used \p{nat} as the
typename. 

The most common combinations of primitive types (\ie basic ACL2
data types)
have been predefined using \p{defdata} (in
``cgen/base'').\footnote{The command \p{(table-alist
    'defdata::types-info-table (w state))} shows the list of all
  types and associated metadata.} There is also a defdata type
called \p{all} which represents the entire ACL2 universe. Every
defdata type is thus a subtype (subset) of \p{all}. Now that we
have a ``base'', of typenames, we can proceed to use them to
build new types.

\paragraph{Product Types}
In ACL2, the primary way to define (complex) structured data is
to use the primitive \emph{data constructor} \p{cons}.

Suppose that as part of modeling a filesystem, we want to define
the type \p{file} that consists of an inode and content, modeled
by a string. We can compound these types using \p{cons}, to
encode a file.

\begin{verbatim}
(defdata file (cons inode string))
\end{verbatim}


\paragraph{Union types}
Very often, one wants to define a type predicate which is a
disjunction of other predicates. Continuing with our example, let's say
we defined a function that returns either well-formed files or the
constant \p{nil}, signifying an error. To specify the return type of
such a function using \p{defdata}, we use the built-in type combinator
\emph{oneof}.
\begin{alltt}
(defdata er-file (oneof file nil))
\end{alltt}

This definition also illustrates an important feature of defdata:
\textbf{singleton types}. Quoted objects and objects that normally
evaluate to themselves, such as \texttt{nil}, represent types that
contain only that single object.

\textbf{Note:} The members or constituents of the union type
expression need not be disjoint. In this regard the \p{oneof}
combinator is closer to the untyped view of things, rather than the
traditional ``sum'' type which is usually a disjoint union. In the
above example, however, the constituents, \p{file} and \p{nil} are
disjoint, \ie the objects satisfying \p{file} are distinct from
\p{nil}.

Union and product type definitions can be arbitrarily nested. Here is
a contrived example:

\begin{alltt}
(defdata UPNest 
  (oneof (oneof (cons (oneof 11 7) pos-list) 'ok symbol-alist) 
         (cons symbol (complex integer -1))
         (oneof (oneof 42 (cons pos file) er-file) t 7/6)
         "nice"))
\end{alltt}
In the above example, notice the use of constructors \p{cons} and
\p{complex}, primitive typenames \p{pos} and \p{symbol}, basic
typenames \p{symbol-alist,pos-list} and \p{integer} (which are
available in the ground theory), previously defined typenames
\p{file} and \p{er-file} and constants (singleton types) 11, 7,
\p{'ok} etc.

\paragraph{Recursive types}
Recursive (or inductive) type expressions involve the \p{oneof}
combinator and product combinations, where additionally there is
a (recursive) reference to the typename being defined.  In general,
well-formed recursive types have a particular form:
\[(\text{\p{defdata}} \,\;M\; (\text{\p{oneof} }b_1 \ldots b_m \; r_1 \dots r_n) \]
where $b_i$ are base type expressions containing only references
to existing defdata types and $r_i$ are recursive type
expressions that contain a reference to $M$ inside a product type
expression. As an example, \p{integer-list} and \p{symbol-alist}
can be defined as follows.

\begin{alltt}
(defdata loi (oneof nil (cons integer loi)))
(defdata symb-alist (oneof nil (cons (cons symbol all) symb-alist)))
\end{alltt}

Mutually-recursive types are also supported. As an example, we
can specify the structure of a directory in a filesystem as
follows; a \emph{dir} is a list of named dir-entries and a
\emph{dir-entry} is either a file or a directory.

\begin{alltt}
(defdata 
  (dir (oneof nil (cons (cons string dir-entry) dir)))
  (dir-entry (oneof file dir)))
\end{alltt}

\paragraph{Range types}
Range types are such a common occurrence that \p{defdata}
supports these natively.\footnote{These are a recent addition and
  are implemented using the infrastructure provided by the
  Tau-system.} To define a range types, we need to provide a
domain, lower and upper bounds, and inequality
relations. We only support \p{integer} and \p{rational}
domains. Both \p{<,<=} are allowed as inequality relations. 
One of the lower or upper bounds can be omitted, in which case the
corresponding value is taken to be negative or positive infinity
(with a strict relation). 

The following two examples show how to define the rational
interval $(0..1)$ and the integers greater than $2^{32}$, say in
order to use the Cgen framework~\cite{acl2-11} to test
exceptional cases involving numbers that do not fit into a 32 bit
machine word.

\begin{alltt}
(defdata cache-miss-ratio (range rational (0 < _ < 1))) 
(defdata big-unsigned-num (range integer ((expt 2 32) < _)))
\end{alltt}

The constructs introduced so far form the \emph{core defdata}
language. Now we look at some convenient type combinations that could
be expressed using the core defdata constructs. These additional
constructs thus can be seen as syntactic sugar. The
motivation for these is not mere typing convenience; some of these
``syntactic sugar'' constructs capture commonly occurring data
definition patterns, and we take the opportunity to automate the
corresponding reasoning patterns.

\paragraph{Macros}
 Macros are freely allowed in the body of \p{defdata}. The meaning is
 what you get after macro expansion.
\begin{verbatim}
(defdata 3d-point (cons rational (cons rational (cons rational nil))))
\end{verbatim}
The above can be defined equivalently and more concisely using the
\p{list} macro.
\begin{verbatim}
(defdata 3d-point (list rational rational rational))
\end{verbatim}

\paragraph{List types}
Nil-terminated lists are so common that we reserve a special
combinator, \p{listof}, for defining list types. Here is how we define
a list of files:
\begin{alltt}
(defdata files (listof file))
\end{alltt}

The List type combinator is the quintessential polymorphic type
operator in typed functional programming languages; moreover due to
their ubiquitous presence in ACL2, list type definitions are subjected
to some automation, \eg a number of theorems are generated to make
reasoning about the recently introduced list type as automated as
possible.

Association lists are also very common in ACL2. It is easy to define
an alist type using \p{cons} and \p{listof}; however, we plan to
introduce an \p{alistof} combinator, for the same reasons we
introduced \p{listof}.
\begin{alltt}
(defdata symbol-alist2 (listof (cons symbol all)))
\end{alltt}

\paragraph{Enumeration types}
If your type is a finite list of ACL2 data objects, then the \p{enum}
keyword can be used to define enumerated types.  Let us define a
subset of opcode instructions, from the MIPS ISA, that use immediate
offsets.
\begin{verbatim}
(defdata opcode (enum '(lw sw addi subi beqz jr)))
\end{verbatim}
Notice that we could have just used the \p{oneof} combinator to
achieve the same effect. But the \p{enum} keyword adds much
convenience; instead of enumerating all the data objects, one can
specify an expression which will evaluate to the finite list of
enumerated objects.  For the above example, here is an alternative
definition.
\begin{verbatim}
(defdata opcode (enum (generate-instruction-opcodes 'mips-risc-model)))
\end{verbatim}
This last specification style is particularly handy if the number of
enumerated objects is very large.

\paragraph{Record (Struct) types}

Records are just product data, but the convenience and ease of use
obtained from named fields deserves special treatment.  We can define
a MIPS instruction as a record, consisting of an opcode, destination
and source register numbers, and the immediate value fields (16 bits).

\begin{alltt}
(defdata reg-num (range integer (0 <= _ < 32)))
(defdata immediate-range (range integer (0 <= _ < (expt 2 16))))
(defdata inst (record (op  . opcode)
                      (rd  . reg-num)
                      (rs1 . reg-num)
                      (imm . immediate-range)))
\end{alltt}


One can also define recursive records; an illustrative example
appears in Section~\ref{sec:full-defdata}.

\paragraph{Map types}

Maps are objects representing finite partial functions. They can be encoded
using alists, but their ubiquity and utility motivates us to treat
them specially. For example, we can define instruction memory to be a map from
physical addresses to instructions:
\begin{alltt}
(defdata p-addr (range integer (0 <= _ < (expt 2 32))))
(defdata imem (map p-addr inst))
\end{alltt}

The advantage of defdata's map types over alist types is that the
underlying implementation guarantees that maps are sorted and have no
duplicate entries; this is quite useful when we are generating
instruction memories for testing purposes. The implementation details
(semantics) of both record and map types are given in
Section~\ref{sec:full-defdata}. But we cannot defer the explanation of
how objects of these types are \emph{constructed} and
\emph{destructed}.

\paragraph{Using record and map data objects}
 For the \p{inst} definition, a 4-arity constructor, \p{inst},
 accessors \p{inst-op}, \p{inst-rd}, \p{inst-rs1}, \p{inst-imm}, and
 modifiers \p{set-inst-op}, \p{set-inst-rd},\p{set-inst-rs1},
 \p{set-inst-imm} are generated.  
 For maps, the accessor and modifier functionality is provided via
 functions \p{mget,mset}. We briefly note here that for both records
 and maps, we make available a useful theory for reasoning about
 compositions of these functions (constructor, accessor, modifier
 etc).

The following code illustrates how to use these functions.

\begin{alltt}
(let* (;; generate a "random" imemory using imem's enumerator
       (I (nth-imem 834546))
       ;; fix a program counter value
       (pc 1)
       ;; get the instruction pointed to by pc
       (instr (mget pc I)) 
       ;; get the immediate value field of instr
       (im (inst-imm instr))
       ;; set the immediate value field and the pc entry 
       (I1 (mset pc (set-inst-imm (1+ im) instr) I))
       ;; an alternative way of getting the immediate value field
       (im2 (mget :imm (mget pc I))))
  ...)
\end{alltt}

\textbf{Note:} The \p{listof,enum,record,map} combinators cannot be
arbitrarily nested and have strict syntax restrictions. In some
instances, the limitation is due to our lack of support for expressing
anonymous recursive types and anonymous functions; in others, it was a
cost-benefit design decision.

\paragraph{Custom Types}
Sophisticated users may want to define types that are more complex
than a union or product combination of existing types. We call such
types \emph{custom types}, and allow the user to define them
manually by providing details that \p{defdata} would generate automatically.
Once a custom type is defined, it can be used just like any other
type to define new types.

For example, suppose you would like to define an instruction memory
but would like the physical addresses to be uniformly ordered from
some address down to 0 . We could define a custom enumerator and
predicate for that purpose:

\begin{verbatim}
(defun make-descending-addresses (n)
  (if (zp n)
      nil
    (cons (1- n) (make-descending-addresses (- n 1)))))

(defun nth-imem-custom (n) ;enumerator
  (let* ((m (nth-imem n))
         (vals (strip-cdrs m))
         (keys (make-descending-addresses (len m))))
    (pairlis$ keys vals)))

(defun imem-customp (x) ;recognizer
  (or (null x)
      (and (consp x) (consp (car x))
           (imem-customp (cdr x))
           (instp (cdar x))
           (p-addrp (caar x))
           (or (and (null (cdr x)) (equal 0 (caar x)))
               (> (caar x) (caadr x))))))
\end{verbatim}

We can now register our custom instruction memory type:

\begin{alltt}
(register-type imem-custom
               :predicate imem-customp
               :enumerator nth-imem-custom)
\end{alltt}

\textbf{Advanced Note:} Instead of defining a new type and polluting
the type name space, we could alternatively have ``attached'' our
custom enumerator to the existing type \p{imem}, using the following
form. This arranges for \p{imem} to be sampled/tested (by Cgen) using
the custom enumerator we defined above, but for theorem proving
purposes, the logical predicate definition of \p{imem} is as
before. This situation can be compared with the \p{:mbe} paradigm,
where one can ``attach'' different logical and execution behaviors to
a function name.\footnote{Here the word ``attach'' should not be
  confused with the \p{defattach} command in ACL2.}
\begin{alltt}
(defdata-attach imem :test-enumerator nth-imem-custom)
\end{alltt}

\section{Defdata Language}
\label{sec:full-defdata}
In this section we will present the syntax and semantics of the
defdata language, \ie defdata type expressions used in the body of the
\p{defdata} form. To do this precisely, we need to explain two
additional macros, \p{register-type}, \p{register-data-constructor}.

\subsection{Registering a type}

We previously saw an example of how to register any custom type as a
\emph{defdata type}, using the \p{register-type} macro. We describe
its syntax below.

\begin{alltt}
(register-type \textit{name}
               :predicate \textit{pred}
               :enumerator \textit{enum}
               \textit{optional args})

(defun nth-odd (n) 
  (if (evenp n)
      (1+ n)
    (- n)))

(register-type odd
               :predicate oddp 
               :enumerator nth-odd)
\end{alltt}

Odd numbers are a basic data type available in the ACL2 ground
theory. They are registered in ``cgen/base.lisp'' as shown
above. 
The predicate and enumerator
arguments are mandatory; the rest are optional.\footnote{We do not
  expose these extra arguments now to the user, as they might change
  in the future.}

This macro, apart from storing relevant metadata in a table (in the
ACL2 world), maintains the following invariants. 
\begin{enumerate}
\item The predicate name is a 1-arity predicate function identified by
  the Tau-system \ie it must have an entry in the Tau-database.\footnote{An exception is the type \p{all}.}
\item The enumerator is a 1-arity function that takes a natural number
  and returns a value of the correct type.\footnote{This check is
    skipped currently. In the current implementation, we also maintain
    an additional 2-arity enumerator function.}
\end{enumerate}

Let \CH{} represent the command history of the current ACL2 session.
We say symbol $d$ is a \emph{registered type name} in history \CH{} if
there exists a command of the form (\p{register-type} $d\; \ldots$)
in \CH.

\subsection{Registering a data constructor}

We have seen various examples of forming compound data types
using \p{cons}. Although \p{cons} has a unique status in ACL2, it
is not natively available in the \p{defdata} language unlike
built-in combinators such as \p{oneof} and \p{range}. In fact,
advanced users can introduce custom notions of product data by
using the \p{register-data-constructor} macro, whose usage and
semantics we now present.

Consider the \p{symbol-alist} type. We could have registered
\p{acons} as a data constructor, and alternatively defined 
\p{symbol-alist} using \p{acons}.

\begin{alltt}
(defun aconsp (x)
  (and (consp x) (consp (car x))))

(register-data-constructor (aconsp acons)
                           ((allp caar) (allp cdar) (allp cdr)))

(defdata symb-alist (oneof nil (acons symbol all symb-alist)))
\end{alltt}

In fact, this is how we setup the base environment in
``cgen/base.lisp'': we use \p{register-data-constructor} to
preregister all the primitive data constructors in ACL2. In
particular, the following (primitive) constructors are available to
build product types: \p{cons}, \p{intern\$}, \p{/} and \p{complex}.

The syntax of \p{register-data-constructor} is shown below.

\begin{alltt}
(register-data-constructor (\textit{recognizer} \textit{constructor})
                           ((\textit{destructor-pred1} \textit{destructor1}) ...)
                           [:proper \textit{bool}]
                           [:hints \textit{hints}]
                           [:rule-classes \textit{rule-classes}])
\end{alltt}

We now explain its semantics. A (\p{register-data-constructor} ($R\;$
$C$) $((D_1\; d_1) \ldots (D_n\; d_n))$) command axiomatizes (checks)
certain properties of the recognizer function $R$, the $n$-ary
constructor $C$, the $n$ destructors (selectors) $d_i$ and the
corresponding destructor predicates $D_i$. In particular it generates
the following properties as \p{defthm} events.

\begin{flalign*}
\big[ \bigwedge_j (D_j\; x_j) \big] &\Ra (R \;\, (C \; x_1 \ldots x_n)) &\text{(Recognizer)}\\
\text{for each $i$}\quad (R\; x) &\Ra (D_i\; (d_i\; x)) &\text{(Destructor predicate)}\\
\intertext{If \texttt{:proper} is true (this is the default), then
  the following two properties are also generated.} \\
(R\;x) &\Ra x = (C\; (d_1\, x) \ldots (d_n\, x))&\text{(Elim [proper])}\\
\text{for each $i$ }\quad \big[ \bigwedge_j (D_j\; x_j) \big] &\Ra x_i = (d_i\; (C\; x_1 \ldots x_n))) &\text{(Destructing a constructor [proper])}\\
\end{flalign*}

Usually, these properties already exist in the ground ACL2 theory as
theorems with appropriate rule-classes for the primitive data
constructors, so for these we default to \p{:rule-classes nil}.
 
Finally, a \p{register-data-constructor} command also stores relevant
metadata (\eg pairs the constructor with its destructors), to be used
in particular by the \p{defdata} implementation.

We say symbol $C$ is a \emph{registered data constructor} in history
\CH{} if there exists a command of the form 
(\p{register-data-constructor} ($R\; \;C$) \dots)  in \CH.

\textbf{Note:} Although \p{acons} is not a primitive data
constructor, because it uses \p{cons}, we nevertheless register
it. We implement record types in a similar manner (explained
later). The ACL2 logic does not allow us to truly define a new
constructor, unlike in NQTHM which provided this capability via
\p{add-shell}. However, by using cons trees and hiding the
internal implementation, we \emph{pretend} to define new data
constructors. As in NQTHM, we would like for all our constructors to
be disjoint with each other:
\begin{flalign*}
 \text{for distinct $C,K$:} \quad(C\; x_1 \ldots x_n) &\neq (K\; y_1 \ldots y_m) &\text{(disjoint C,K)}
\end{flalign*}
In ACL2 this is true for all the primitive constructors, but we
cannot always enforce this property for the reasons explained above.
We do enforce this property to the extent that we can, \eg when
we implement records, we tag a unique name to \emph{constructed}
objects so that the objects constructed from different record
types are disjoint with one another, if not with \p{cons}.

\subsection{Core Defdata - Syntax and Semantics}
Now we present the core defdata language; in particular, we describe
the syntax and semantics of the core type expressions used in the body
of a defdata form. The syntax of a \p{defdata} form is as follows:
\begin{alltt}
(defdata \(M\) \textit{type-expression})
\end{alltt}
For a mutually-recursive clique of types, we use a syntax similar to \p{defuns}:
\begin{alltt}
(defdata (\(M\sb{1}\) \textit{type-expression\(\sb{1}\)}) \dots)
\end{alltt}

We now explain the syntax of a core defdata type expression. In the
following, we use syntactic (meta) variables $s,t$ to range over type
expressions, $a,b,c$ to range over constant symbols and also to range
over objects of the ACL2 universe, $P,Q$ to range over monadic
predicates and $A,B$ to range over typenames, both primitive/custom
types registered by \p{register-type} and typenames introduced by
\p{defdata} itself. Let $M$ be the name of a defdata type being
defined. To avoid introducing a recursive binding operator ($\mu$) and
type variables to express anonymous recursive types, we assume below
that $M$ is registered. We have opted for clarity over completeness,
so the syntax we present is abstract; our implementation performs
additional syntactic checks not mentioned here.

\begin{flalign*} 
\intertext{{\it (Type expressions)}}
t ::=&\;\;\;\p{c}       &\text{(quoted) constant}\\
     &\B A        &\text{registered type name}\\
     &\B \all     &\text{top}\\
&\B (\textsf{range}\;\, \mathit{dom}\;\, r) \quad  \mathit{dom} \in \{\textrm{integer},\textrm{rational}\} &\text{range}\\
     &\B  (\oneof \,\; t_1 \ldots t_m) \quad m > 1 &\text{union}\\
     &\B  (C \;t_1 \ldots t_n) &\text{} \\
&\quad C \mbox{ is a registered
  constructor with } ((D_1\; d_1) \ldots (D_n\;d_n)) \text{ and } \bigwedge_i \subtype{t_i}{D_i} \hspace*{-130pt} & \text{product}\\
     &\B (\oneof\; s_1 \ldots s_m\;\; t_1 \ldots t_n) \quad m>0,n>0 &\text{}\\ 
& \quad \textit{ $t_i$ is product, $M \in t_i, M\not\in s_j$, where $\in$ is} \textrm{ ``occurs in''} &\text{recursive}
\intertext{{\it (range expressions)}}
r ::= \quad & (l\ \p{< \_ <}\ h)  \B (l\ \p{< \_ <=}\ h) \B
(l\ \p{<= \_ <}\ h) \B (l\ \p{<= \_ <=}\ h) &\text{$l,h$ eval to objects in $\mathit{dom}$}\\
         \B & (\p{\_ <}\ h) \B (\p{\_ <=}\ h) &\text{negative infinity}\\
         \B & (l\ \p{< \_})\; \B (l\ \p{<= \_}) &\text{positive infinity}\\
\end{flalign*}

The subtype relation, $\subtype{}{}$, is the subset relation among
the set of objects satisfying the type expressions; its precise meaning will
be given shortly. 

We now touch upon the semantics of core defdata type expressions and
of the \p{defdata} command.  What happens when a (\p{defdata} \p{M}
 $s$) event is submitted is not easy to capture neatly, due to the fact
that it has been engineered over many years to satisfy sometimes very
different goals between specification convenience and test data
generation (Cgen) efficacy. Nevertheless, we try to give a reasonably
good model of what happens, and hope that future refactorings and
design changes do not render our explanation obsolete.

\renewcommand{\P}{\mathcal{P}}
\newcommand{\E}{\mathcal{E}}

Apart from syntax checking, there are, broadly, four things that a
core defdata command (\p{defdata} \p{M} $\,s$) accomplishes.
\begin{enumerate}
\item Introduces a predicate definition event (\p{defun Mp} (x)
  $\;\P(s)(x)$) (or a defuns clique, in case of a mutually-recursive
  type definition), if \p{Mp} is not defined. If the predicate named
  \p{Mp} is already defined, it checks the equivalence of the new and
  old definitions.
\item Introduces an enumerator definition event (\p{defun}
  \p{nth-M} (n) $\;\E(s)(n)$). For infinite domains, we would
  ideally like for this function to be a bijection from the natural
  numbers to the domain of $M$. At the very least we would like its
  range to be adequate for Cgen. The
  current implementation also defines a second enumerator
  function of arity 2, whose second argument is a random
  seed that is threaded through nested enumerator
  calls.\footnote{We might add/remove alternative enumerative
    characterizations of a type in the future; but we are
    confident we will maintain at least one such
    characterization.}
\item Introduces rules (\p{:tau-system} and others) that capture the
  type relations induced by the command between the defined type $M$
  and the typenames in $s$ (see
  Section~\ref{sec:tau-characterize} for details). 
\item For use by subsequent calls to defdata, it registers the
  typename $M$ with its predicate and enumerator
  names and records other relevant metadata.
\end{enumerate}

We now give the predicative characterization of type
expressions. Each core defdata type expression denotes a subset
of the ACL2 universe and is characterized by a predicate lambda
expression. The predicate interpretation $\P$ shows how to 
compile type expressions to ACL2s code. Given a type expression,
$\P$ generates a lambda expression in ACL2 that takes one
argument and returns either \p{t} or \p{nil}.


\begin{flalign*}
\P(\p{a}) =& \;\lambda x.(x = \p{a}) \\
\P(A) =& \;\lambda x.(Q\; x) &\text{$A$ is registered with predicate name $Q$}\\
\P(\textsf{all}) =& \; \lambda x. \p{t} &\text{symbol \p{t} in ACL2 stands for $\true$}\\
\P((\oneof\; t_1 \ldots t_m)) =& \; \lambda x. \bigvee_i^m \P(t_i)(x) \\
\P((C\; t_1 \ldots t_n)) =& \;\lambda x. (R\;x) \And
\bigwedge_i^n \P(t_i)(d_i\; x) \hspace*{-75pt} &\text{$C$ is registered with recog $R$ and dest $d_i$}
\end{flalign*}
We add the definitional event (\p{defun Mp} $(x)\; \P(s)(x)$), assuming
$M$ is registered when computing $\P(s)$.  When generating code for
(mutually-)recursive types, $\P$ generates (mutually-)recursive
definitions. Such definitions are not necessarily well-defined. ACL2s
uses its CCG termination analysis engine~\cite{ManoliosVroon2006} to
check for termination and only accepts a defdata form if CCG can prove
termination.

\textbf{Note:} The subtype relation among type expressions stands for
the inclusion relation between their predicate interpretations. We also
abused notation (in the syntax of product type expressions) using
predicate $D_i$ as a type expression, instead of using the typename of
$D_i$.
\[ \mathfrak{I}(t \sqsubseteq s) := \; \P(t) \Ra \P(s) \]
Now we turn to the enumerative characterization of type
expressions. By the same reasoning as before, each core defdata
type expression can be characterized by some enumerator function
on \(\mathbb{N}\). The enumerator interpretation $\E$ takes a
type expression and generates an ACL2 lambda expression that take
a natural number as an argument and returns an object of the
right type.
\begin{flalign*}
\E(\p{a}) =& \;\lambda n.\p{a} \\
\E(A) =& \;\lambda n.(E^A\; n) &\text{$A$ is registered with enumerator  } E^A\\
\E(\textsf{all}) =& \; \lambda n. (\p{nth-all }n)
&\text{\p{nth-all} enumerates the ACL2 universe}\\
\E((\oneof\; t_1 \ldots t_m)) =& \; \lambda n. (\textsf{mv-let}\;
(i\;\; n')\; (\mathit{switch}\;m\;n)\;\;\;
\E(t_i)(n')) \hspace*{-35pt} & \\
\E((C\; t_1 \ldots t_k)) =& \;\lambda n.  (\textsf{mv-let}\; (n_1
\ldots\, n_k)\;\; (\mathit{split}\;k\;n)
(C\;\;\E(t_1)(n_1)\ldots\, \E(t_k)(n_k))) \hspace*{-180pt} & 
\end{flalign*}
We add the definitional event (\p{defun nth-M} $(n)\; \E(s)(n)$),
assuming $M$ is registered when computing $\E(s)$.  Notice that the
definition of \p{nth-M} generated by $\E$ might be recursive. As
before, ACL2s depends on the CCG termination analysis engine to prove
that such definitions make sense.

The helper functions \emph{switch} and \emph{split}\footnote{The
  actual functions are named \p{defdata::switch-nat} and
  \p{defdata::split-nat}.}, are used in defining enumerator
expressions for union and product type expressions respectively.
We refer to $n$ above as an \emph{indicial}, an index into the
type domain.  Given $m$ choices and indicial $n$, the expression
\((\mathit{switch}\;m\;n)\) returns $i$, a number between 0 and $m-1$
denoting which type to ``switch'' to, and $n'$, a new 
indicial to pass on. Given number $k$ and indicial $n$, the
expression \((\mathit{split}\;m\;n)\) ``splits'' $n$ into $k$
indicials. Both functions are designed to be bijective. Thus, if
the constituent types have bijective enumerators, a valuable
meta-property, then \emph{switch} and \emph{split}, preserve that
property for the type combination. Not all primitive and basic
enumerators defined and registered in ``cgen/base.lisp'' are
bijective; in particular \p{nth-all} only heuristically
enumerates (interesting portions of) the ACL2 universe. 

The Cgen library uses a pseudo-geometric random distribution to
generate (usually small) indicials, which are used to randomly sample
test data. For nested product types, the split indicials obtained
after multiple levels of splitting usually end up as a bunch of 0s;
this is natural, \(\lambda n . (\mathit{split}\; k\; n)\) is a
bijection between natural numbers and $k$-tuples of natural
numbers. This skews the test data generation for complex product
types. To get around this, we also generate a more complex,
accumulator-based enumerative characterization very similar to the one
documented above.  Instead of one argument, the enumerator function
carries a (pseudo-random) seed as a second argument, and
threads it through the sequence of enumerator calls. This results in a
more uniform distribution of test data for product types. Instead of
\emph{switch}, it uses the 2-arity \p{random-index} function, that
takes numbers $\mathit{m}$ and $\mathit{seed}$ and returns a number
between 0 and $m-1$ and a new random seed. It avoids \emph{split}
altogether by using the 2-arity \p{random-natural} function directly,
that returns a random indicial and a new seed.

We use the above semantics of type expressions to mechanically
generate the predicate and enumerator functions for each type defined
using \p{defdata}.\footnote{Range type expressions currently cannot be
  nested and are implemented using \p{make-tau-interval} and
  \p{in-tau-intervalp} and enumerator functions for \p{rational} and
  \p{integer}.} The Cgen library can be set to use either of the
enumerators.

\subsection{Full defdata language}

We will now fill in the rest of the combinators that \p{defdata}
supports. We only briefly discuss macros and enum types. Macro names
that occur in the position of a combinator or constructor are expanded
away using the function \p{macroexpand1}. Enum types, though
expressible using \p{oneof} combinator, are treated natively; in
(\p{defdata} $M$ (\p{enum} \textit{list-expr})) the \textit{list-expr}
is evaluated and defined separately as a defconst, that is then used
to define the predicate and enumerator.


\paragraph{List types}
For list data definitions we have the following expansion, where $s$
can be any core defdata type expression.
\[(\p{defdata } M\; (\p{listof } s)) = \; (\p{defdata } M\;\;(\p{oneof} \;\p{nil  } (\p{cons } s\;\; M)))\]

Whereas this suffices to take care of the four things a core
defdata command accomplishes (as per the previous section), the
\p{listof} combinator does a lot more. In particular, \p{defdata}
installs some useful list reasoning theorems that are commonly
needed in a proof development using lists. It also performs some
processing to support polymorphic list type reasoning, whose
discussion we postpone to Section~\ref{sec:poly}.

\paragraph{Record types}
Record data definitions are simulated by a combination of (core) defdata commands and 
\p{register-data-constructor}.  Record type
expressions have named fields; their syntax is as follows:
\\ \p{(record (f1 . t1) \dots (fk . tk))} where \p{f1,\dots,fk} are
symbols (field names) and \p{t1,\dots,tk} are typenames. One can also
have records that have a fresh constructor name \p{(C (f1 . t1)
  \dots)}; these are usually combined with the \p{oneof}
combinator.\footnote{Readers familiar with ML, will notice the
  similarity to the syntax of \textit{datatype} facility.}  We will
explain their semantics using an example. The general case can be
easily extrapolated. Let us define a binary tree, as a (recursive)
record, with non-leaves having three fields, \p{val} storing data
associated with that node, \p{left} for the left subtree and \p{right}
for the right subtree.
\begin{alltt}
(defdata tree (oneof 'Leaf 
                     (node (val   . all)
                           (left  . tree)
                           (right . tree))))
\end{alltt}

Let us assume that \p{node} is a fresh logical name. The above
definition can be expanded to the following form that is almost in the
core defdata language.

\begin{alltt}
(defdata tree (oneof 'Leaf 
                     (node all tree tree)))
\end{alltt}

This is ``almost'' in the language because \p{node} is not a
registered data constructor so \p{defdata} cannot generate the
predicate and enumerator functions for \p{tree}. We need to get
our hands on at least two things, a constructor and the
accessors. It is natural to use \p{node} as the constructor name
and to use the field names in the accessor names.  To avoid
name-clashes, we use \p{node-val}, \p{node-left} and
\p{node-right} as the accessor/destructor function names. For the
above (core) defdata form to have meaning, we first register
\p{node} as a new constructor.

\small
\begin{alltt}
(register-data-constructor (nodep node)
                           ((allp node-val) (treep node-left) (treep node-right)))
\end{alltt}
\normalsize

But there is a chicken-and-egg problem. To submit this command, we
need the predicates \p{nodep} and \p{treep} to be defined and to generate
\p{nodep, treep}, we need this command to be in the history. We get
around this by assuming the metadata that a
\p{register-data-constructor} command usually records, generating the
predicate definitions first and then submitting the above command.

There is still the issue of how to define the constructors and
accessors, \ie how do we implement the layout of the record? Efficient
reasoning of compositions of these functions motivates our decision to
implement records as ``good''
maps~\cite{ManoliosKaufmann02,02-kaufmann-rewriting,DBLP:journals/jfp/GreveKMMRRSVW08}.
A good-map is an ordered alist with non-nil value components (see
definition in file:
\texttt{defexec/other-apps/records/records}).\footnote{Implementation
  note (subject to change): Each record of $k$ fields is implemented
  as a good-map containing $k+1$ entries. The extra entry
  ('DEFDATA::CONSTRUCTOR . name) stores the name of the constructor
  and thus allows easy disjointedness (with other records)
  theorems. Each field \textit{sel} has the corresponding entry with
  key \p{:sel} and thus is accessed by the expression \p{(mget :sel
    r)}, where \p{r} is the record object.}

Non-recursive records use a special keyword \p{record}, as we saw in
the examples. For the above semantics to apply to it, we need to come
up with a name for the constructor; \p{defdata} reuses the name of the
type being defined, \eg the \p{inst} data definition, we saw earlier,
is equivalent to the following.

\begin{alltt}
(defdata inst (inst (op . opcode) ...))
\end{alltt}


\paragraph{Map types}
Map data definitions, or finite functions types, are expanded using
the following equation, where $s,t$ are restricted to be defdata type
names.
\[(\p{defdata } M\; (\p{map}\;\; s\;\; t)) = \; (\p{defdata } M \;\;(\p{oneof} \;\p{nil}\;\; (\p{mset } s\;\;t\;\; M)))\]

For map types, the \p{defdata} macro directly reuses the implementation
of \p{good-map}~\cite{DBLP:journals/jfp/GreveKMMRRSVW08}. The
constructor \p{mset} is registered in \texttt{cgen/base}; here is the
relevant excerpt:\footnote{The guard declarations have been removed for
  readability.}

\begin{alltt}
(defun non-empty-good-map (x)
  (and (consp x)
       (good-map x)))

(defun all-but-nilp (x)
  (not (equal x 'nil)))

(register-data-constructor (non-empty-good-map mset)
                           ((wf-keyp caar) (all-but-nilp cdar) (good-map cdr)))
\end{alltt}

Both record and map definitions additionally introduce useful theorems
that help in termination proofs and type-like reasoning (in
particular involving constructor, accessor and updater functions).

\section{Characterizing type relations induced by defdata}
\label{sec:tau-characterize}
A number of queries to the ACL2 theorem prover, especially in
guard verification, involve establishing inclusion (subtyping)
among types (monadic predicates) and proving that certain terms
satisfy given types. This sort of type-checking has received a
boost in automation by the addition of Tau-system to the ACL2
proof arsenal~\cite{enhancementsKM13}. Due to the impossibility
of automatically inferring relations among arbitrarily defined
recursive predicates, it is up to the user to inform the
Tau-database by stating theorems (called tau-rules) describing
the subtype relationships and function signatures. In a perfectly
informed Tau-database, the Tau-system can, in theory, turn into a
``complete'' procedure, type-checking all queries in its domain
correctly and automatically. In this section, we describe how
\p{defdata} programs Tau to maintain this goal of (relative)
completeness.

Given a core defdata definition \p{(defdata M $s$)}, we would
like to compute the set of tau rules (see \p{:doc tau-system})
that completely characterize the inclusion/exclusion type
relationship between $M$ and typenames in $s$.\footnote{At the
  time of writing, the given scheme has only been partially
  implemented.} This is very useful as it leads to a more
systematic and profitable use of the Tau-system, enabling
automated type-like reasoning. With such a scheme in effect, one
need not worry about manually determining and proving all
relationships between the newly defined type and those used in
its definition. Thus, if the type relations among base (and
custom) types are completely captured in Tau, then extending the
Tau-database by types specified using \p{defdata} preserves this
meta-property (under a suitable restriction on the form of the
types).

After the defdata form \p{(defdata M $s$)} is successfully
admitted, a predicate $P = \P(s)$ is defined.  The idea
is quite simple: we decompose the definition into two
implications and reduce each implication into  a
collection of tau rules. If we show that each reduction scheme is
sound and complete then the final set of formulas will completely
characterize the type relations induced by the defdata command.
{\vskip 10pt} \newcommand{\CategoryB}[1]{\noindent \textbf{#1:}}
\CategoryB{$\P(s)(x)$ $\Ra$ $(P\; x)$}

Let $C_1,\ldots, C_m$ be the conjunctive clauses of the disjunctive
normal form (DNF) of $\P(s)(x)$.

\[ \infer*[right=(elim OR)]{C_1 \Ra (P\;x) \\ \cdots \\  C_m \Ra (P\;x) }{\P(s)(x) \Ra (P\;x)} \]
The above reduction scheme is clearly an equivalent transformation.

The following scheme performs one level of destructor-elimination to
reduce destructor nesting in the antecedent ($C_i$) in exchange for
constructor nesting in the succedent. The expression $(\mathbf{Q}\;x)$
either stands for $(Q\; x)$ (where $Q$ is a tau predicate) or another
destructor nest \((R' x) \wedge (\mathbf{Q}'_1\;(d'_1 x)) \wedge
\cdots \wedge (\mathbf{Q}'_n (d'_n\;x))\). We say the \emph{head} of
$\mathbf{Q}$ is $Q$ in the former and $R'$ in the latter case.

\[ \infer*[right=(dest elim)]{\{C \text{ is registered with } R\;\; d_1\ldots d_n\}  \\ (\mathbf{Q_1} \;x_1) \wedge \cdots \wedge (\mathbf{Q_n} \;x_n) \Ra (P\;(C\; x_1 \ldots x_n)) }{(R\;x) \wedge (\mathbf{Q_1} \;(d_1\;x)) \wedge \cdots \wedge (\mathbf{Q_n}\; (d_n\;x)) \Ra  (P\;x) } \]

This transformation is sound and complete, because syntactically valid
defdata product expressions satisfy the type signature of the
constructor, \ie the head of $\mathbf{Q_i}$ implies (is a subtype of)
$D_i$, where $D_i$ is the corresponding destructor predicate for $d_i$
in $C$.  Therefore adding/dropping $(R\;x)$ does not change the
truth-value. 

We can only apply these reductions finitely often. After doing so
we obtain implications that are either simple rules, signature
rules, or neither. Any implications that are simple or signature
rules can be turned into tau rules. All other implications 
have to consist of nested constructor calls; such implications
cannot be directly turned into tau rules. 

Consider the example of \p{files}, whose definition has been
expanded to a core defdata expression.  \small
\begin{alltt}
\(\P\)(\p{(oneof nil (cons file files))})(x) \(\Ra\) (filesp x)
\quad \(\longrightarrow\) \{\emph{Def. of \(\P\), DNF form}\}
(or (= nil x) (and (consp x) (filep (car x)) (filesp (cdr x)))) \(\Rightarrow\) (filesp x)
\quad \(\longrightarrow\) \{\(\textsc{Elim OR}\)\}
(= nil x) \(\Rightarrow\) (filesp x) [Simple Rule]
(and (consp x) (filep (car x)) (filesp (cdr x))) \(\Rightarrow\) (filesp x)
\quad \(\longrightarrow\) \{\(\textsc{Dest Elim}\)\}
(and (filep x1) (filesp x2)) \(\Rightarrow\) (filesp (cons x1 x2)) [Signature Rule]
\end{alltt}
\normalsize
Next, consider the definition of \p{symb-alist}, which gives rise to a
formula with nested constructor calls, \ie it does not conform to a tau
rule. We will use $P$ for \p{symb-alistp}.
\small
\begin{alltt}
\(\P\)(\p{(oneof nil (cons (cons symbol all) M))})(x) \(\Ra\) (P x)
\quad \(\longrightarrow\) \{\emph{Def. of \(\P\), DNF form}\}
(or (= nil x) (and (consp x) (consp (car x)) (symbolp (caar x)) (P (cdr x))) \(\Rightarrow\) (P x)
\quad \(\longrightarrow\) \{\(\textsc{Elim OR}\)\}
(= nil x) \(\Rightarrow\) (P x) [Simple Rule]
(and (consp x) (and (consp (car x)) (symbolp (caar x)) (P (cdr x))) \(\Rightarrow\) (P x)
\quad \(\longrightarrow\) \{2 applications of \(\textsc{Dest Elim}\)\}
(and (symbol x12) (P x2)) \(\Rightarrow\) (P (cons (cons x11 x12) x2)) [Not a tau rule]
\end{alltt}
\normalsize




We say $s$ is a \emph{flat type expression}, if product type
expressions in $s$ have only typename arguments. Similarly, if $s$ is
flat, we say that \p{(defdata M $s$)} is a \emph{flat definition} and
$M$ is a \emph{flat type}, \eg \p{files} is a flat type, but
\p{symb-alist} is not. Since flat types lead to only valid tau rules,
in this direction, we obtain a characterization of the type relations
induced by \p{defdata} in terms of tau rules.

In general, the problem of nested constructor calls can be taken
care of either by (1) extending the Tau-system to handle such
cases, or by (2) introducing intermediate data definitions that
name nested union and product combinations and thus getting rid
of the nesting (\ie making the definition flat). The latter
scheme fails when a recursive reference to the typename is nested
more than one level deep in a product expression. But this is not
common, and we feel flat definitions are a suitable restriction
that covers the majority of data definitions of interest and
utility.

{\vskip 10pt}

\CategoryB{$(P\; x)$ $\Ra$ $\P(s)(x)$} 

In the other direction, we can symmetrically try the dual approach:
let $D_1, \ldots, D_m$ be the disjunctive clauses of the conjunctive
normal form (CNF) of $\P(s)(x)$.

\[ \infer*[right=(elim AND)]{(P\;x) \Ra D_1 \\ \cdots \\  (P\;x) \Ra D_m}{(P\;x) \Ra \P(s)(x)} \]
The above reduction scheme is an equivalent transformation too. For
flat definitions with an additional restriction of having at most one
occurrence of a product type expression, one can check that the final
(irreducible) formulas will be valid tau rules and we complete the
characterization in both directions.

But the above scheme precludes some useful data definitions, such as
those with two recursive product expressions; in this case even
intermediate naming does not help. We are currently working on an
alternative approach that avoids this difficulty.






\section{Supporting Polymorphism}
\label{sec:poly}

Polymorphic reasoning can significantly increase automation.
Consider a typical example. A user of ACL2s has proved some
rewrite rules about lists of files (\p{filesp}), but the rules
are not firing as expected. After some investigation the user
discovers the problem: they did not explicitly prove that
\p{append} is closed over lists of files, hence, ACL2s was not
able to determine that \p{filesp} holds for \p{(append x y)},
even though \p{x} and \p{y} satisfy \p{filesp}. The solution is
simple: the user has to prove that \p{append} is closed over
lists of files. But, why should the user have to do that? After
all, this is really a property of \p{append}, not
\p{filesp}. Polymorphic reasoning solves this problem.  The idea
is that the user tags certain theorems mentioning \p{true-listp}
in their conclusion as ``polymorphic'' and ACL2s will treat such
theorems as schemas that hold for all predicates of the same
shape. All useful examples of such polymorphic theorems, that we
are aware of, are type signatures of (polymorphic) functions. So
instead of modifying the syntax of defthm-like events and
defdata, we need only provide a syntax extension to defun-like
forms (\eg \p{defunc}, \p{define}, etc.) to allow polymorphic
type signatures. We also need to change the semantics of defdata
events to provide the invariant that all instances of the
polymorphic type signatures are present in the current
theory. These two changes suffice to simulate a form of
``parametric polymorphism'' in ACL2s.

\subsection{Expressing polymorphic signatures}
\newcommand{\sig}{\p{sig}}

The polymorphic support in ACL2s depends on encapsulation and
functional instantiation. We use macros to hide this from the
end-user.  The \sig{} macro expresses polymorphic signatures. In the
future, we would like to integrate it with the \p{defunc} macro.
The syntax and usage of \sig{} is best explained by examples.
\begin{verbatim}
(sig nthcdr  (nat (listof :a)) => (listof :a))

(sig zip  ((listof :a) (listof :b)) => (listof (cons :a :b)))

(sig assoc-equal  (:a (listof (cons :a :b))) => (oneof nil (cons :a :b)))

(sig binary-append  ((listof :a) (listof :b)) => (listof (oneof :a :b)))

General Form:
(sig fun-name  arg-types => return-type)
\end{verbatim}

Type variables are represented by keyword symbols, \p{:a, :b, \dots}
and types of arguments are given using defdata type expressions, with
special handling of keyword symbols (type variables).


We show by example how the semantics of \sig{} is implemented in ACL2s.



 




\begin{verbatim}
(sig binary-append  ((listof :a) (listof :b)) => (listof (oneof :a :b)))
==> 
\end{verbatim}

\begin{verbatim}
(encapsulate 
 (((Ap *) => *) ((Bp *) => *))
 
 (local (defun Ap (v)
          (declare (ignore v))
          t))

 (local (defun Bp (v)
          (declare (ignore v))
          t))

 (defthm Ap-is-predicate
   (booleanp (Ap x)))

 (defthm Bp-is-predicate
   (booleanp (Bp x))))

(defun LoAp (xs)
  (if (endp xs)
      t
    (and (Ap (car xs))
         (LoAp (cdr xs)))))

(defun LoBp (xs)
  (if (endp xs)
      t
    (and (Bp (car xs))
         (LoBp (cdr xs)))))

(defun LoCp (xs)
  (if (endp xs)
      t
    (and (or (Ap (car xs)) (Bp (car xs)))
         (LoCp (cdr xs)))))

 (defthm binary-append-polymorphic-sig
   (implies (and (LoAp x)
                 (LoBp y))
            (LoCp (binary-append x y)))))
\end{verbatim}

The names of constrained functions are chosen appropriately and we
reuse existing names if possible. The predicate bodies are generated
using the predicate interpretation of type expressions given in
Section~\ref{sec:full-defdata}.

\subsection{Putting polymorphism to use (behind the scenes)}

So we can express polymorphic type signatures, but how do we make
use of them? The answer is via functional instantiation. We want
to hide this from the user for pedagogical and usability
reasons. We accommodate this by ensuring the following:
\begin{enumerate}
\item Every time the user introduces a new defdata type that
  is an instance of a parameterized type used in a polymorphic type
  signature, we immediately use functional instantiation to submit the
  corresponding instantiated type signature for the newly introduced
  type as a rewrite rule. The following example illustrates this.

\begin{verbatim}
(defdata even-list (listof even))
==> 
...
(defthm binary-append-even-listp-sig
  (implies (and (even-listp x)
                (even-listp y))
           (even-listp (binary-append x y)))
  :hints (("Goal"
           :use
           ((:functional-instance
             binary-append-polymorphic-sig
             ;; Instantiate the generic functions:
             (Ap evenp)
             (Bp evenp)
             ;; Instantiate the other relevant functions:
             (LoAp even-listp)
             (LoBp even-listp)
             (LoCp even-listp))))))
...
\end{verbatim}

\item For every \sig{} event, we look into the Tau-database,
  collecting all similar shape instances of the polymorphic type
  expressions used in the signature, that have not already been
  instantiated. For each such instance we introduce the corresponding
  instantiated type signature as a rewrite rule. As an example, as
  soon as the polymorphic signature for \p{binary-append} is
  introduced, the macro also generates the following events (where
  \p{nat-listp}, \p{pos-listp}, \ldots{} are types already present in
  the Tau-database).

\begin{verbatim}
(sig binary-append  ((listof :a) (listof :b)) => (listof (oneof :a :b)))
==>
...
(defthm binary-append-nat-list-sig
  (implies (and (nat-listp x)
                (nat-listp y))
           (nat-listp (binary-append x y)))
  :hints (("Goal"
           :use
           ((:functional-instance
             binary-append-polymorphic-sig
             (Ap natp)
             (Bp natp)
             (LoAp nat-listp)
             (LoBp nat-listp)
             (LoCp nat-listp))))))

(defthm binary-append-pos-list-sig
  (implies (and (pos-listp x)
                (pos-listp y))
           (pos-listp (binary-append x y)))
  :hints (("Goal"
           :use
           ((:functional-instance
             binary-append-polymorphic-sig
             (Ap posp)
             (Bp posp)
             (LoAp pos-listp)
             (LoBp pos-listp)
             (LoCp pos-listp))))))
...
\end{verbatim}




\end{enumerate}

Our approach can wind up generating a lot of events, especially
if there are many signatures with multiple type variables. All of
the rules that we generate in support of polymorphic reasoning
are tau rules, so they are automatically added to the Tau-database by ACL2. A consequence of this is that for datatypes
that are completely characterized by tau rules, we maintain that
completeness even in the presence of polymorphic reasoning.

\section{Related Work}
\label{sec:rel}

There are a number of macro libraries in the ACL2 Community books that
specify data definitions and capture common reasoning patterns. The
oldest of these libraries are \p{defstructure} by Bishop
Brock~\cite{brockdefstructure} and \p{deflist,defalist} by Bill
Bevier. These libraries can be found in the community books
\texttt{data-structures} directory of the ACL2 distribution.

Towards mechanizing the proof of soundness of typed lambda
calculus~\cite{swords2006soundness}, Sol Swords developed the
\p{defsum} macro (found in tools/defsum.lisp) that provides a
convenient syntax for specifying mutually-recursive types.

Jared Davis has contributed a number of useful macros for specifying
typed lists, alists, enums, unions and records that can be found in
\texttt{std/util}~\cite{Davis/util}.

All of these libraries, like \p{defdata} generate a lot of
events, and in particular, install an extensive set of theorems
that automate reasoning about the defined types and functions
operating on them. In fact, we believe these libraries are more
advanced than \p{defdata} with regard to theorem proving
automation. A distinguishing feature is that we are integrated
with the Cgen library and we maintain an enumerative
characterization of the type definitions. 

More recently, Sol Swords has written a macro library
(FTY)~\cite{Swords:fty} for supporting a particular discipline of
using types in ACL2. It associates a fixing function (\eg \p{nfix})
and an equivalence relation with each type predicate in addition to
providing the usual constructs to define mutually-recursive types.






\section{Conclusion and Future Work}
\label{sec:conc}

We presented the \p{defdata} type definition framework. Our framework
provides a convenient mechanism for defining, testing, and reasoning
about datatypes in ACL2s. We have used \p{defdata} to teach about
1,000 undergraduate students at Northeastern University how to reason
about programs. In conjunction with the \p{defunc} macro, which allows
one to define functions with input and output contracts, \p{defdata}
provides type-like capabilities in ACL2s. We provided a partial
characterization of data definitions using Tau, hence, reasoning about
data definitions is highly automated.  We also showed how to support
polymorphic reasoning in ACL2s. Our framework is also used by experts
and is available to regular ACL2 users as a community book.

For future work, we would like to further integrate \p{defunc},
\p{defdata} and Tau for increased automation, efficiency, and
debugging capabilities. We plan to provide a flexible API to the
\p{defdata} framework and to work with the ACL2 community to help
create a standardized data definition framework. We plan to provide
support for more advanced forms of data definitions such as dependent
types, quotient types, predicate subtypes, and intersection types.

\section*{Acknowledgments}
This research was supported in part by DARPA under AFRL
Cooperative Agreement No.~FA8750-10-2-0233 and by NSF grants
CCF-1117184 and CCF-1319580.

\bibliographystyle{eptcs} \bibliography{paper}
\end{document}